\documentclass[12pt]{article}
\title{ On the notion of a macroscopic quantum system}
\author{Andrei Khrennikov\\International Center for Mathematical
Modeling \\ in Physics and Cognitive Sciences,\\
University of V\"axj\"o, S-35195, Sweden\\
Email:Andrei.Khrennikov@msi.vxu.se}
\begin{document}
\maketitle

\abstract{It is proposed to define "quantumness" of a system (micro or macroscopic,
physical, biological, social, political) by starting with
understanding that quantum mechanics is a statistical theory. It
says us only about probability distributions. The only possible
criteria of quantum behaviour are statistical ones. Therefore I
propose to consider any system which produces quantum statistics as
quantum ("quantumlike"). A possible test is based on
the interference of probabilities. I was mainly interested in using
such an approach to "quantumness" to extend the domain of
applications of quantum mathematical formalism and especially to
apply it to cognitive sciences. There were done experiments on interference of
probabilities for ensembles of students and  a nontrivial
interference was really found. One could say that the quantum statistical behaviour
might be expected. But the problem was not so trivial. Yes, we might
expect nonclassical statistics, but there was no reason to get the
quantum one, i.e., cos-interference. But we got it!}

\bigskip

The notion of a "macroscopic quantum system" might play an important
role in better understanding of quantum mechanics and domains of its
application. During my discussion with Antony Leggett last summer
(conference on "Quantum and Mesoscopic Dynamics") it became evident
for me that in fact the notion a macroscopic quantum system is not
defined in a rigor way. Different people have rather different
views. Typically a macroscopic quantum system is understood as a
macroscopic physical system which is able to stay in a
"superposition" of two different states. However, even for
microscopic systems we are not able to observe superposition of
states for an individual system. In the microdomain this problem can
be ignored, since here "micro" is typically used as a synonymous of
"quantum". In the macrodomain we could not ignore this problem.

I would like to pay attention to the fact that (at least for me) it is not completely clear:

\medskip

{\it ``What can be called a macroscopic quantum system?''}

\medskip

Of course, this question is closely related to the old question:

\medskip

{\it ``What can be called a quantum system?''}

\medskip

There is no common point of view to such notions as quantization,
quantum theory.  For me (in the opposition to N. Bohr) the presence of quanta (of, e.g., energy)
is not the main distinguishing feature of quantum theory. Of course, the presence of observables (e.g., energy) with discrete spectra is an important feature of  quantum theory. However, the basic quantum observables, the position and the momentum, still have continuous ranges of values. I think that the main point is that quantum theory is a {\it statistical theory.} Therefore it should be characterized in statistical terms. We should find the basic feature of quantum theory which distinguishes this theory form classical statistical mechanics.  The {\it interference of probabilities} is such a basic statistical feature of quantum theory.
Therefore any system (material or not) which exhibits (for some observables) the interference
of probabilities should be
considered as a quantum system (or say ``quantum-like system'').\footnote{The terminology ``quantum theory''
is rather misleading. It may be better to call it ``theory of probabilistic interference'',
instead of  quantum theory?}
 Thus, since human beings by replying to special pairs of questions produce
interference of probabilities, they should be considered as macroscopic quantum systems.\footnote{Experiment to test interference of probabilities for complementary
questions to people was designed in [1], see also [2], and it was performed and experimental
statistics demonstrated the interference of probabilities [3]. In [1], [2] there was used so called
contextual approach to statistical measurements. This is classical (but contextual) probabilistic
approach which produces in particular the representation of probabilistic data by complex probability
amplitudes (or in the abstract form in the complex Hilbert space), see [4]--[6] for details.}

 I presented
this viewpoint to macroscopic quantum systems in my discussions with A. Leggett (after his public lecture
in Prague connected with the Conference ``Frontiers of Quantum and Mesoscopic Thermodynamics'', Prague, July-2004,
and  during my talk at University of Illinois). Unfortunately, neither A. Leggett nor other participants
of the conference buy my idea on human being as a macroscopic quantum system.\footnote{I emphasize that
my approach has nothing to do with quantum physical reduction of mental processes which was intensively discussed
in a lot of publications, see, e.g., Penrose [7]. In my approach human being is statistically
quantum as a complex information system. This quantum statistical behavior could not be reduced
to quantum statistical behavior
of photons, electrons,... composing the human body.}

As I understood, for physicists a macroscopic quantum system is
a huge ensemble of microscopic quantum systems (e.g., electrons) prepared in a special state.
Human being is also a huge ensemble of microscopic quantum systems... However, the state of this
ensemble cannot be considered as quantum from the traditional point of view. Nevertheless, according to
our interference viewpoint to quantumness human being is quantum (but not because it is composed of
microscopic quantum systems).

In the connection with our discussion on the definition of a macroscopic quantum system it is natural
to mention experiments of A. Zeilinger and his collaborators [8] on interference of probabilities for
fullerens and other macromolecules including biomolecule porphyrine. It seems that A. Zeilinger uses the same definition of quantumness as I. It is interesting that one of the main aims of further experiments of A. Zeilinger is to
find the interference of probabilities for some viruses. I think that at that point he will come really very close
to my viewpoint to macroscopic quantum systems, in particular, biological quantum systems.

\medskip

{\bf References}

1. A. Yu. Khrennikov, On cognitive experiments to test quantum-like behaviour
of mind. {\it Reports from V\"axj\"o University,} N 7 (2002);
http://xxx.lanl.gov/abs/quant-ph/0205092 (2002).

2. Khrennikov A.Yu., {\it Information dynamics in cognitive, psychological and
anomalous phenomena.} Ser. {\it Fundamental Theories of Physics.} Kluwer, Dordreht (2004).

3. E. Conte, O. Todarello, A. Federici, F. Vitiello, M. Lopane, A. Khrennikov,
A preliminar evidence of quantum-like behaviour in measurements of mental states.
Reports from MSI, N. 03090, 2003; Proc. Int. Conf. {\it Quantum Theory: Reconsideration
of Foundations.} Ser. Math. Modelling in Phys., Engin., and Cogn. Sc., vol. 10,
679-702, V\"axj\"o Univ. Press, 2004. quant-ph/0307201.

4. A. Yu. Khrennikov, Interference of probabilities and number field structure of quantum models.
{\it Annalen  der Physik,} {\bf 12}, N. 10, 575-585 (2003).

5. A. Yu. Khrennikov, Contextual approach to quantum mechanics and the theory of the
fundamental prespace. {\it J. Math. Phys.}, {\bf 45}, N.3, 902-921 (2004). quant-ph/0306003.

6. A. Yu. Khrennikov, E. Loubenets, On relation between probabilities in quantum and classical experiments. {\it Foundations of  Physics,} {\bf 34,}, N 4, 689-704(2004).

7. R. Penrose, {\it The emperor's new mind.} (Oxford Univ. Press, New-York, 1989).

R. Penrose  {\it Shadows of the mind.}  (Oxford Univ. Press, Oxford, 1994).

8. A. Zeilinger, Exploring the boundary between the quantum and classical worlds.
{\it  Int. Conf. ``Frontiers of Quantum and Mesoscopic Thermodynamis''},  book of abstracts,
p. 66, Prague, July-2004.

\end{document}